\documentclass[10pt,twocolumn,superscriptaddress,longbibliography,prx]{revtex4-2}

\usepackage{amsmath, bm, amsfonts, amssymb}     % maths symbols
\usepackage{graphicx}                       % coloured text, subfigs
\usepackage{xfrac}                          % inline fractions
\usepackage{grffile}                        % filenames with dots
\usepackage{color,colortbl}
\usepackage{xcolor}
\usepackage{multirow}

\newcommand{\Gin}{G_{\rm in}}
\newcommand{\Gc}{G_{\rm c}}
\newcommand{\Go}{G_{\rm out}}
\newcommand{\Gcone}{G_{\rm c1}}
\newcommand{\Gctwo}{G_{\rm c2}}
\newcommand{\Gcthree}{G_{\rm c3}}
\newcommand{\Gcn}{G_{\rm cn}}
\newcommand{\nuin}{\nu_{\rm in}}

\newcommand{\nuo}{\nu_{\rm out}}
\newcommand{\nucone}{\nu_{\rm c1}}
\newcommand{\nuctwo}{\nu_{\rm c2}}
\newcommand{\nucthree}{\nu_{\rm c3}}
\newcommand{\nucn}{\nu_{\rm cn}}
\newcommand{\Rin}{R_{\rm in}}
\newcommand{\Rc}{R_{\rm c}}
\newcommand{\dV}{dV}
\newcommand{\de}{d\epsilon}
\newcommand{\dg}{d\gamma}

\newcommand{\rhat}{\boldsymbol{\hat{\textbf{r}}}}
\newcommand{\that}{\boldsymbol{\hat{\theta}}}
\newcommand{\ur}{u_r}
\newcommand{\urp}{u_{rp}}
\newcommand{\urcp}{u_{rcp}}
\newcommand{\ursp}{u_{rsp}}
\newcommand{\ut}{u_\theta}
\newcommand{\utp}{u_{\theta p}}
\newcommand{\utcp}{u_{\theta cp}}
\newcommand{\utsp}{u_{\theta sp}}
\newcommand{\err}{\epsilon_{rr}}
\newcommand{\ett}{\epsilon_{\theta\theta}}
\newcommand{\ert}{\epsilon_{r\theta}}
\newcommand{\srr}{\sigma_{rr}}
\newcommand{\stt}{\sigma_{\theta\theta}}
\newcommand{\srt}{\sigma_{r\theta}}
\newcommand{\bvec}{\boldsymbol{\textbf{b}}}
\newcommand{\uvec}{\boldsymbol{\textbf{u}}}

\newcommand{\atens}{\boldsymbol{\textbf{a}}}

\usepackage[colorlinks, citecolor=blue, urlcolor=red, linkcolor=blue]{hyperref}

\begin{document}

\title{Simple and effective mechanical cloaking}

\author{Suzanne M. Fielding}
\affiliation{Department of Physics, Durham University, Science Laboratories,  South Road, Durham DH1 3LE, UK}

\begin{abstract}

We show theoretically that essentially perfect elastostatic mechanical cloaking of a circular inclusion in a homogeneous surrounding medium can be achieved by means of a simple cloak comprising three concentric annuli, each formed of a  homogeneous isotropic linear elastic material of prescribed shear modulus. Importantly, we find that the same combination of annuli will cloak any possible mode of imposed deformation or loading, for any randomly chosen admixture of imposed compression, pure shear and simple shear,  without the need to re-design the cloak for different deformation modes.  A full range of circular inclusions can be cloaked in this way, from soft  to stiff. In consequence,  we suggest that an inclusion of any arbitrary shape can also be cloaked, by first enveloping it in a stiff circle, then cloaking the combined structure with  three annuli as described. Given that a single inclusion can be fully cloaked in this way, even at near field close to the cloaking perimeter, it also follows that multiple such neutral inclusions arranged with arbitrarily high packing fraction in a surrounding medium can also be cloaked. We confirm this by direct simulation. This indicates a possible route to fabricating composite materials with the same global mechanical response as a counterpart homogeneous material, and with uniform strain and stress fields outwith the cloaked inclusions.

\end{abstract}

\maketitle

Cloaking is the practice of rendering unnoticeable an inclusion that differs  in some physical way from its surrounding medium, by enveloping it in a cloak.  Widespread potential applications include optical invisibility~\cite{pendry2006controlling,leonhardt2006optical}, isolation from electromagnetic fields~\cite{wood2007metamaterials,zhu2015three,gomory2012experimental}, sound reduction in acoustics~\cite{norris2008acoustic,cummer2007one,chen2007acoustic},  drag reduction in hydrodynamics~\cite{park2019hydrodynamic}, thermal protection~\cite{xu2014ultrathin},  mass shielding~\cite{guenneau2013fick},  the mitigation of elastodynamic vibrations~\cite{brun2009achieving,nassar2019isotropic,norris2012hyperelastic,farhat2009ultrabroadband,zhang2020asymmetric}, including for seismic protection~\cite{brule2017flat}, and -- the focus of this work -- the elastostatic cloaking of inclusions in elastic media to achieve  mechanical ``unfeelability''. 

The last two decades have seen rapid progress in cloaking  technologies, as recently reviewed~\cite{martinez2022metamaterials}. In the context of electromagnetism~\cite{pendry2006controlling,leonhardt2006optical}, invariance of the governing field equations under transformation has been exploited to achieve cloaking via the  use of   conformal mapping to engineer a suitably heterogeneous and anisotropic refractive index. In elasticity theory, however, the governing equations are not invariant~\cite{yavari2019nonlinear}: transformation in general leads to an anisotropic density and a coupling between strain and momentum~\cite{milton2006cloaking}. Some successes have nonetheless been achieved in the elastodynamic cloaking of propagating waves, for example by means of a spatially varying elasticity tensor of rank four~\cite{brun2009achieving}, hyperelastic materials under pre-stress~\cite{norris2012hyperelastic} or asymmetric metamaterials~\cite{zhang2020asymmetric}. Flexural waves in thin sheets have also been successfully cloaked~\cite{farhat2009ultrabroadband}. 

In the context of elastostatic cloaking,  the basic aim is to render an inclusion in an elastic medium mechanically ``unfeelable'', with  the deformation, strain and stress fields outside the cloaked inclusion being the same as they would be in the equivalent homogeneous medium without an inclusion. Following Eshelby's seminal work on inclusions~\cite{eshelby1957determination}, the search for neutral inclusions  has a long history~\cite{mansfield1953neutral,christensen1979solutions,sozio2023optimal}, often however involving complicated imperfect interfaces~\cite{bertoldi2007structural,bigoni1998asymptotic,he20023d,ru1998interface,wang2012neutrality}, anistropic  materials~\cite{benveniste1991effective,hashin1990thermoelastic,norris2020static}, precise tuning of both a material's shear modulus and Poisson ratio~\cite{wang2023double,wang2023neutrality}, or considering only antiplane elasticity~\cite{milton2001neutral} or a purely compressional deformation, without shear~\cite{kang2016coated}. Indeed, progress in elastostatic cloaking lags significantly behind that in other fields, not least due to the breakdown of transformation methods noted above. Some success has nonetheless recently 
 been achieved by employing  pentamode metamaterials~\cite{buckmann2014elasto}, direct lattice transformation~\cite{buckmann2015mechanical}, morphable voids~\cite{cheng2023compatible}, shape optimisation~\cite{fachinotti2018optimization}, topology optimization~\cite{ota2022mechanical}, lattice based metamaterials~\cite{sanders2021optimized} and data-driven aperiodic metamaterial design~\cite{wang2022mechanical}.

Typically, however, such strategies involve complicated metamaterial cloaking structures  comprising hundreds or thousands of  subunits, requiring computationally intense optimisation  and challenging fabrication. They furthermore often achieve only modest cloaking performance, as quantified by the measure of field disturbance compared with the affine field in a homogeneous medium, which is typically reduced   only by a factor $O(10)$ or exceptionally $O(100)$ compared with the uncloaked case. More importantly still, while such cloaks can be optimised to achieve reasonable to good performance (as just quantified) for one particular imposed mode of deformation or loading -- shear, compression, or one particular prescribed admixture of both -- a different cloak must be designed  for each admixture of deformations or loads separately. 

In this work, we show by analytical calculation and direct numerical simulation (which agree fully) that essentially perfect mechanical cloaking of a circular inclusion in $d=2$ spatial dimensions can  be achieved by a simple cloak comprising three concentric annuli, each formed of a  homogeneous isotropic linear elastic material of a prescribed elastic modulus, which can be computed in seconds to minutes on a laptop. We  hope that such cloaks will also be relatively straightforward to fabricate experimentally. Importantly  we show that the same combination of annuli will cloak any possible mode of deformation and/or loading, with any randomly chosen admixture of compression, pure shear and simple shear, without the need design a different cloak for each different admixture.  We show that a full range of circular inclusions can be cloaked in this way, from soft  to stiff. This furthermore  suggests that an inclusion of any arbitrary shape can also be cloaked, by first enveloping it a stiff circle, then cloaking the combined structure with three annuli as just described. 

A key feature of our approach is that, for any given combination of inclusion and surrounding material, it suffices only to tune the values of the shear moduli $\Gcn$ in the cloaking annuli, for any set of annuli Poisson ratios $\nucn$. Indeed, we emphasise that these $\nucn$ can  be chosen arbitrarily, prior to then tuning (for any set of $\nucn$) the moduli $\Gcn$ according to the approach that we set out below. This  contrasts with recent work~\cite{wang2023double,wang2023neutrality}, in which both  the shear modulus and the Poisson ratio need to be precisely prescribed in each region of the cloak.  This distinction is important in particular from the viewpoint of  fabricating a device in practice, due to the near impossibility of precisely prescribing both the shear modulus and Poisson ratio for any component material.

The manuscript is structured as follows. In Sec.~\ref{sec:setup} we introduce the cloaking geometry and governing equations. Sec.~\ref{sec:analytics}  uncovers an exact analytical condition for perfect cloaking, which we then confirm by direct numerical simulation in Sec.~\ref{sec:numerics}. We present our concluding perspectives in Sec.~\ref{sec:conclusion}.

\section{Geometry and governing equations}
\label{sec:setup}

\begin{figure}[!t]
\begin{center}
      \includegraphics[width=0.8\columnwidth]{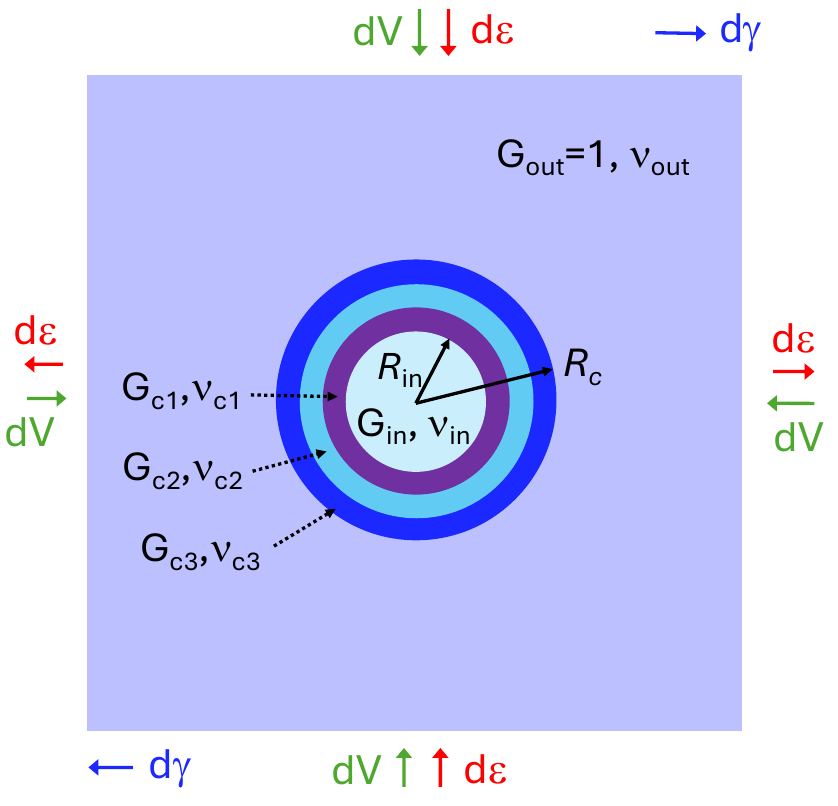} 
\end{center}
  \caption{At the level of linear isotropic elasticity in $d=2$ spatial dimensions, we study a circular inclusion of radius $\Rin$ and elastic modulus $\Gin$ in a surrounding medium of modulus $\Go$, subject to an imposed deformation comprising an arbitrary admixture of bulk isotropic compression  $\partial_x U_x=\partial_y U_y=-\dV$, pure shear $\partial_x U_x=-\partial_y U_y=\de$ and simple shear  $\partial_y U_x=\dg$. Our aim is to surround the inclusion by a cloak extending from inner radius $\Rin$ to outer radius $\Rc$ comprising one, two or three  annuli with  elastic moduli $\Gcn$, such that the deformation field outside the cloak will be undisturbed compared with the affine imposed one that would obtain in a homogeneous medium without an inclusion.} 
  \label{fig:sketch} 
\end{figure}

As sketched in Fig.~\ref{fig:sketch}, at the level of linear isotropic elasticity in $d=2$ spatial dimensions, we consider a circular inclusion of radius $\Rin$, elastic modulus $\Gin$ and Poisson ratio $\nuin$ surrounded by a cloak extending from inner radius $\Rin$ to outer radius $\Rc$  in a surrounding medium  of elastic modulus $\Go$ and Poisson ratio $\nuo$. We  explore in what follows cloaks comprising one, two or three annuli  with moduli $\Gcn$ and Poisson ratios $\nucn$.  When two or three annuli are present, we assume that all annuli in any cloak  are of equal thickness.

The governing equations are as follows. As a function of space $r_{i}$  we define a strain field 
\begin{equation}
\label{eqn:strain}
    \epsilon_{ij}(r_i)=\tfrac{1}{2}\left(\partial_iu_j+\partial_ju_i\right)
\end{equation}
in terms of a displacement field  $u_i(r_i)$ relative to  undeformed equilibrium.
 At the level of linear isotropic elasticity, the elastic stress field
\begin{equation}
\label{eqn:stress1}
\sigma_{ij}(r_i)=2\mu\epsilon_{ij}+\lambda\epsilon_{ll}\delta_{ij}.
\end{equation}
Force balance in the inertialess limit requires that
\begin{equation}
\label{eqn:forceBalance}
0_i=\partial_j\sigma_{ij}.
\end{equation}

In Eqn.~\ref{eqn:stress1}, the constants $\mu$ and $\lambda$ are the material's Lam\'e coefficients, which can be expressed in terms of its shear and bulk moduli $G$ and $K$ by the relations
\begin{equation}
\label{eqn:LameElastic}
    \mu=G\;\;{\rm and}\;\; \lambda=-\frac{2G}{d}+K.
\end{equation}
The material's Poisson ratio
\begin{equation}
\label{eqn:Poisson}
\nu=\frac{dK-2G}{d(d-1)K+2G}.
\end{equation}
Among these five constants $\mu,\lambda,G,K$ and $\nu$ we need to specify only two,  because the other three are then determined by relations~\ref{eqn:LameElastic} and~\ref{eqn:Poisson}. We shall work in terms of $G$ and $\nu$, noting that in $d=2$ dimensions the ratio  $f\equiv\lambda/\mu=2\nu/(1-\nu)$. 

A uniform affine deformation field comprising some  admixture of bulk isotropic compression $\partial_x U_x=\partial_y U_y=-\dV$ (or expansion, $dV<0$), pure shear $\partial_x U_x=-\partial_y U_y=\de$ and  simple shear $\partial_y U_x=\dg$ is imposed at far field $r\to\infty$ in our analytical  calculation in Sec.~\ref{sec:analytics} and on average across a biperiodic box of size $L\times L$ in our direct numerical simulations in Sec.~\ref{sec:numerics}. We use upper case $U_i$ to denote this affine part of the deformation field, distinct from the full field $u_i$, which in general deviates from it and depends on space $r_i$.

We choose as our stress unit the modulus of the surrounding medium  $\Go=1$. Throughout most of what follows we choose as our length unit the inclusion radius $\Rin=1$, except in Fig.~\ref{fig:snapshot}, where we simulate multiple inclusions in the same box. No time unit is needed, because we  consider  static cloaking.

\section{Analytical condition for perfect cloaking}
\label{sec:analytics}

In this section, we derive an analytical condition for perfect mechanical cloaking in the limit of linear isotropic elasticity. We assume that the interfaces between inclusion, cloaking annuli and surrounding medium are all perfectly bonded and sharp.  (In our simulations in Sec.~\ref{sec:numerics},  each interface will have a small non-zero thickness set by the numerical grid. This will give a small correction to perfect cloaking, which nonetheless decreases as the inverse  linear grid density.)

Our tactic  will be to obtain first a solution for the deformation field in the inclusion, in each annulus and in the surrounding medium for any arbitrarily chosen set of annular moduli  $\Gcn$ and Poisson ratios $\nucn$, not initially aimed at cloaking. This solution will have a deformation field in the surrounding medium that in general differs from the affine field in a homogeneous medium. We then use this general uncloaked solution as a springboard to investigate what conditions must apply to the annular moduli $\Gcn$ such that the non-affine part of the deformation field outside the cloak is eliminated, giving perfect cloaking.

\subsection{General solution without cloaking}

\subsubsection{Bulk solution}
\label{sec:bulk}

In cylindrical polar coordinates centred on a circular inclusion, the deformation field
\begin{equation}
    \uvec=\ur(r,\theta)\rhat+\ut(r,\theta)\that.
\end{equation}
The strain field (Eqn.~\ref{eqn:strain}) is then given componentwise as
\begin{eqnarray}
\err&=&\partial_r\ur,\nonumber\\
\ert&=&\tfrac{1}{2}\left(\partial_r\ut+\tfrac{1}{r}\partial_\theta\ur-\tfrac{\ut}{r}\right),\nonumber\\
\ett&=&\tfrac{1}{r}\partial_\theta \ut +\tfrac{\ur}{r},
\end{eqnarray}
and the stress field (Eqn.~\ref{eqn:stress1}) as
\begin{eqnarray}
\label{eqn:stressComp}
\srr&=&2G\,\partial_r\ur+Gf\left(\partial_r\ur+\tfrac{1}{r}\partial_\theta\ut+\tfrac{\ur}{r}\right),\nonumber\\
\srt&=&G\left(\partial_r\ut+\tfrac{1}{r}\partial_\theta\ur-\tfrac{\ut}{r}\right),\nonumber\\
\stt&=&2G\left( \tfrac{1}{r}\partial_\theta \ut +\tfrac{\ur}{r} \right)+Gf\left(\partial_r\ur+\tfrac{1}{r}\partial_\theta\ut+\tfrac{\ur}{r}\right).
\end{eqnarray}
We recall here the ratio $f=\lambda/\mu=2\nu/(1-\nu)$ defined above.
%, with $\mu=G$.

The condition of force balance (Eqn.~\ref{eqn:forceBalance}) reads:
\begin{eqnarray}
\label{eqn:forceComp}
0&=&\partial_r\srr+\tfrac{1}{r}\partial_\theta\srt+\tfrac{1}{r}(\srr-\stt),\nonumber\\
0&=&\partial_r\srt+\tfrac{1}{r}\stt+\tfrac{2}{r}\srt.
\end{eqnarray}

A system with a cloak comprising $C$ annuli has $C+2$ bulk regions  of uniform modulus $G$ and Poisson ratio $\nu$: the inclusion, the $C$ cloaking annuli, and the surrounding medium.  Combining Eqns.~\ref{eqn:stressComp} and~\ref{eqn:forceComp} gives two equations governing  the deformation field in each such homogeneous region:
\begin{eqnarray}
\label{eqn:bulkEqns}
0&=&2\partial_r^2\ur+\tfrac{1}{r}\partial_\theta \left(\partial_r\ut+\tfrac{1}{r}\partial_\theta\ur-\tfrac{\ut}{r}\right)\nonumber\\
 & &+ \tfrac{2}{r}\left(\partial_r\ur-\tfrac{1}{r}\partial_\theta \ut-\tfrac{\ur}{r}\right) + f\partial_r\left(\partial_r\ur+\tfrac{1}{r}\partial_\theta\ut+\tfrac{\ur}{r}\right),\nonumber\\
 0&=& \partial_r\left(\partial_r\ut+\tfrac{1}{r}\partial_\theta\ur-\tfrac{\ut}{r}\right)+\frac{2}{r}\partial_\theta\left(\tfrac{1}{r}\partial_\theta\ut+\tfrac{\ur}{r}\right)\nonumber\\
 & & \tfrac{2}{r}\left(\partial_r\ut+\tfrac{1}{r}\partial_\theta\ur-\tfrac{\ut}{r}\right)+\tfrac{f}{r}\left(\partial_r\ur+\tfrac{1}{r}\partial_\theta\ut+\tfrac{\ur}{r}\right).
\end{eqnarray}
The solution of these in any  homogeneous bulk region is 
\begin{eqnarray}
\label{eqn:bulk}
    \ut=\sum_{p=-1,1}\utp r^p + \sum_{p=-3,-1}^{1,3}\left[\utcp  \cos(2\theta) + \utsp  \sin(2\theta)\right]r^p,\nonumber\\
     \ur=\sum_{p=-1,1}\urp r^p + \sum_{p=-3,-1}^{1,3}\left[\urcp  \cos(2\theta) + \ursp  \sin(2\theta)\right]r^p,\nonumber\\
\end{eqnarray}
with constant coefficients $\utp,\urp$ for $p=-1,1$ and $\utcp,\utsp,\urcp,\ursp$ for $p=-3,-1,1,3$ calculated in each bulk region separately. In each bulk region, this solution is subject to the following 8 constraints among the coefficients:
\begin{equation}
%\label{eqn:constraint1}
    u_{rc3}=\alpha_3 u_{\theta s3},\;\; u_{rc1}=\alpha_1 u_{\theta s1},
    \end{equation}
    \begin{equation}
%    \label{eqn:constraint2}
    u_{rc-1}=\alpha_{-1} u_{\theta s-1},\;\; u_{rc-3}=\alpha_{-3} u_{\theta s-3},
\end{equation}
\begin{equation}
%\label{eqn:constraint3}
    u_{rs3}=-\alpha_3 u_{\theta c3},\;\; u_{rs1}=-\alpha_1 u_{\theta c1},
    \end{equation}
    \begin{equation}
    \label{eqn:constraint4}
    u_{rs-1}=-\alpha_{-1} u_{\theta c-1},\;\; u_{rs-3}=-\alpha_{-3} u_{\theta c-3},
\end{equation}
in which 
\begin{equation}
\alpha_3=-f/(6+4f),\;\;\alpha_1=-1,\;\;\alpha_{-1}=-2+f,\;\;\alpha_{-3}=1.
\end{equation}

We thus  have 12 unknown coefficients in each bulk region: four at the zeroth angular mode $u_{r1},u_{r-1},u_{\theta 1},u_{\theta -1}$, four at the $\sin(2\theta)$ mode $u_{\theta s3},u_{\theta s1},u_{\theta s-1},u_{\theta s-3}$ and four at the $\cos(2\theta)$ mode $u_{\theta c3},u_{\theta c1},u_{\theta c-1},u_{\theta c-3}$. This gives $12(C+2)$ coefficients in total in our system of $C+2$ bulk regions.

We recognise however that the deformation field  must tend to zero as $r \to 0$. This gives 6 further constraints  within the inclusion: $u_{r-1}=u_{\theta -1}=u_{\theta s-1}=u_{\theta s-3}=u_{\theta c-1}=u_{\theta c-3}=0$. It must likewise tend to the affine imposed one as $r\to \infty$. This gives 6 further constraints  in the surrounding medium: $u_{r1}=V,u_{\theta 1}=-\gamma/2,u_{\theta s3}=0,u_{\theta s1}=-\epsilon,u_{\theta c3}=0$ and $u_{\theta c1}=\gamma/2$.  This reduces the number of unknown coefficients to 
%$12(C+2)-6-6=12(C+1)$. 
$12(C+1)$.
To summarise, these are  
$u_{r1},u_{\theta 1}$ with $u_{\theta s3},u_{\theta s1}$ and $u_{\theta c3},u_{\theta c1}$ in the inclusion;   $u_{r-1},u_{\theta -1}$ with $u_{\theta s-1},u_{\theta s-3}$ and $u_{\theta c-1},u_{\theta c-3}$ in the surrounding medium; and  $u_{r1},u_{r-1},u_{\theta 1},u_{\theta -1}$ with $u_{\theta s3},u_{\theta s1},u_{\theta s-1},u_{\theta s-3}$ and $u_{\theta c3},u_{\theta c1},u_{\theta c-1},u_{\theta c-3}$ in each of the $C$ annuli.  

\subsubsection{Interface conditions}

So far, we have found a solution for the bulk deformation field expressed in terms of the $12(C+1)$  coefficients just listed. To determine the values of these coefficients, given any combination of values of the elastic moduli  and Poisson ratios of the inclusion $\Gin,\nuin$, annuli $\Gcn,\nucn$ and surrounding medium $\Go=1,\nuo$, we consider now the  conditions that apply across the $(C+1)$ interfaces that separate the inclusion and inner cloak annulus, each of the cloaking annuli, and the outer cloak annulus from the surrounding medium. 

We assume that each interface is perfectly bonded and perfectly sharp, with continuous deformation and traction across it: 
%\begin{equation}
$\left[\ur\right]=\left[\ut\right]=\left[\srr\right]=\left[\srt\right]=0$,
%\end{equation}
%
where $\left[c\right]$ denotes the jump in any quantity $c$ across an interface.  For an interface at radius $R$ these conditions read:
\begin{eqnarray}
0&=&\left[\ur\right],\label{eqn:interface1}\\
0&=&\left[\ut\right],\label{eqn:interface2}\\
0&=&\left[(2+f)G\partial_r\ur\right]+\tfrac{1}{R}(\partial_\theta \ut+\ur)\left[fG\right],\label{eqn:interface3}\\
0&=&\left[G\partial_r\ut\right]+\tfrac{1}{R}(\partial_\theta \ur-\ut)\left[G\right].\label{eqn:interface4}
\end{eqnarray}
Across each of our $C+1$ interfaces, each of these 4 constraints must apply separately to the zeroth angular mode, the $\sin(2\theta)$ mode and the $\cos(2\theta)$ mode. This gives $12(C+1)$ linear equations governing the $12(C+1)$ coefficients listed at the end of Sec.~\ref{sec:bulk}. These can be solved numerically in $O(1s)$ on a laptop. Substituting the computed coefficients into Eqn.~\ref{eqn:bulk} then gives the deformation field in all bulk regions: the inclusion, each cloaking annulus and surrounding medium.

\subsection{Conditions for perfect cloaking}

In any system, the bulk deformation field just computed will depend on the elastic moduli and Poisson ratios in the inclusion $\Gin,\nuin$ and annuli $\Gcn,\nucn$ relative to that in the surrounding medium $\Go=1,\nuo$. Indeed, for any arbitrarily chosen $\Gin,\nuin$ and $\Gcn,\nucn$, the deformation field in the surrounding medium in general deviates from the affine one that would  obtain in a homogeneous medium without an inclusion. This deviation is quantified by the coefficients  $u_{r-1},u_{\theta -1},u_{\theta s-1},u_{\theta s-3},u_{\theta c-1},u_{\theta c-3}$  (the other coefficients $u_{r s-1}$ {\it etc.} being specified in terms of these via Eqn.~\ref{eqn:constraint4}). We therefore define the metric
\begin{equation}
\label{eqn:ga}
g=|u_{r-1}|+|u_{\theta -1}|+|u_{\theta s-1}|+|u_{\theta s-3}|+|u_{\theta c-1}|+|u_{\theta c-3}|.
\end{equation}
Perfect cloaking will be attained if this metric can be reduced to zero in the surrounding medium, such that the deformation field outside the cloak reduces to the affine one.

To consider how we might achieve this, it is helpful first to recognise that the $12(C+1)$ interfacial equations  in the previous subsection decouple into four subsystems:

I) At zeroth angular mode, the interfacial conditions~\ref{eqn:interface2} and~\ref{eqn:interface4} on $\ut$ and $\srt$ give $2(C+1)$ linear equations, which we write in generalised matrix form as $\atens\cdot\uvec=\bvec$, in the $2(C+1)$ unknowns $\uvec$ comprising $u_{\theta 1}$ in the inclusion, $u_{\theta -1}$ in the surrounding medium and $u_{\theta 1},u_{\theta -1}$ in each annulus.  The vector $\bvec$ has only one non-zero entry, equal to the imposed shear $u_{\theta 1}=-\gamma/2$ in the surrounding medium. This corresponds to solid body rotation, which is a global invariant, so this subsystem does  not  need considering further. Indeed, the condition $|u_{\theta 1}|=0$ in the inclusion, noted above, automatically ensures $|u_{\theta -1}|=0$ in the surrounding medium. 

II) At zeroth angular mode, the interfacial conditions~\ref{eqn:interface1} and~\ref{eqn:interface3} on $\ur$ and $\srr$ give $2(C+1)$ equations $\atens\cdot\uvec=\bvec$ (redefining $\atens,\bvec$ and $\uvec$ from above) in the $2(C+1)$ unknowns $\uvec$ comprising $u_{r 1}$ in the inclusion, $u_{r -1}$ in the surrounding medium and $u_{r 1},u_{r -1}$ in each annulus. The vector $\bvec$ has only one non-zero entry: the imposed bulk compression  $u_{r1}=\dV$ in the surrounding medium.

III) The interfacial conditions~\ref{eqn:interface1} and~\ref{eqn:interface3} on $\ur$ and $\srr$ at the $\cos(2\theta)$ mode, and~\ref{eqn:interface2} and~\ref{eqn:interface4}  on $\ut$ and $\srt$ at $\sin(2\theta)$, give  $4(C+1)$ equations $\atens\cdot\uvec=\bvec$ (again redefining $\atens,\bvec$ and $\uvec$) in the $4(C+1)$ unknowns $\uvec$ comprising $u_{\theta s3},u_{\theta s1}$ in the inclusion, $u_{\theta s-1},u_{\theta s-3}$ in the surrounding medium and  $u_{\theta s3},u_{\theta s1},u_{\theta s-1},u_{\theta s-3}$ in each  annulus. The vector $\bvec$ has only one non-zero entry: the imposed pure shear $u_{\theta s1}=-\de$ in the surrounding medium.

IV) The interfacial conditions~\ref{eqn:interface2} and~\ref{eqn:interface4}  on $\ut$ and $\srt$ in the $\cos(2\theta)$ mode, and~\ref{eqn:interface1} and~\ref{eqn:interface3}  on $\ur$ and $\srr$ in the $\sin(2\theta)$ mode, give  $4(C+1)$ equations  $\atens\cdot\uvec=\bvec$ in the $4(C+1)$ unknowns $\uvec$ comprising $u_{\theta c3},u_{\theta c1}$ in the inclusion, $u_{\theta c-1},u_{\theta c-3}$ in the surrounding medium and  $u_{\theta c3},u_{\theta c1},u_{\theta c-1},u_{\theta c-3}$ in each annulus. The matrix $\atens$ here is in fact the same as for sub-system III). Likewise the vector $\bvec$ has only one non-zero entry, in the same location as for III), equal to the imposed simple shear $u_{\theta c1}=\gamma/2$ in the surrounding medium.  Indeed, any simple shear can be expressed as the sum of a solid body rotation and a pure shear. The rotation was accounted for in subsystem I). The component here in IV) is pure shear. Accordingly, subsystems III) and IV) are in effect the same: III) describes the pure shear $\de$ with horizontal and vertical principal axes and IV) the pure shear component of $\dg$ with diagonal principal axes.

In summary: of the three components of any general imposed deformation, the compression $\dV$ is governed by subsystem II) and the simple and pure shear  $\dg$ and $\de$ by subsystem IV)=III).  With this in mind, we now proceed to uncover analytical conditions for perfect mechanical cloaking. As a pedagogical warmup discussion, we shall consider first in Sec.~\ref{sec:compression} cloaking compression alone and then (separately) in Sec.~\ref{sec:shear} shear alone. Finally in Sec.~\ref{sec:mixed}, we shall consider the problem of  primary interest: cloaking arbitrary admixtures of  compression, pure shear and simple shear.

For definiteness and simplicity, we shall  perform our numerics to confirm these analytical predictions initially assuming a uniform Poisson ratio $\nu=\nuin=\nucn=\nuo$ across the entire structure, comprising the inclusion, all cloaking annuli and the surrounding material. Indeed, we consider a wide range of values of $\nu$, from conventional to auxetic. We emphasise, however, that our analytical approach does not depend on the Poisson ratio either being uniform across the device, or even (for example) different but precisely tuned in each cloaking annulus. As noted above, this point is important from the viewpoint of  fabricating a device experimentally, because  prescribing both the shear modulus and Poisson ratio of any given material component of a device would be almost impossible to achieve in practice. We shall then return in Sec.~\ref{sec:conclusion} to perform numerics confirming that, for any given inclusion and surrounding material, the Poisson ratios of the cloaking rings can be chosen essentially arbitrarily, with only the shear moduli of the cloaking rings then needing to be precisely prescribed, given any set of arbitrarily chosen Poisson ratios.

\begin{figure}[!t]
\begin{center}
      \includegraphics[width=0.9\columnwidth]{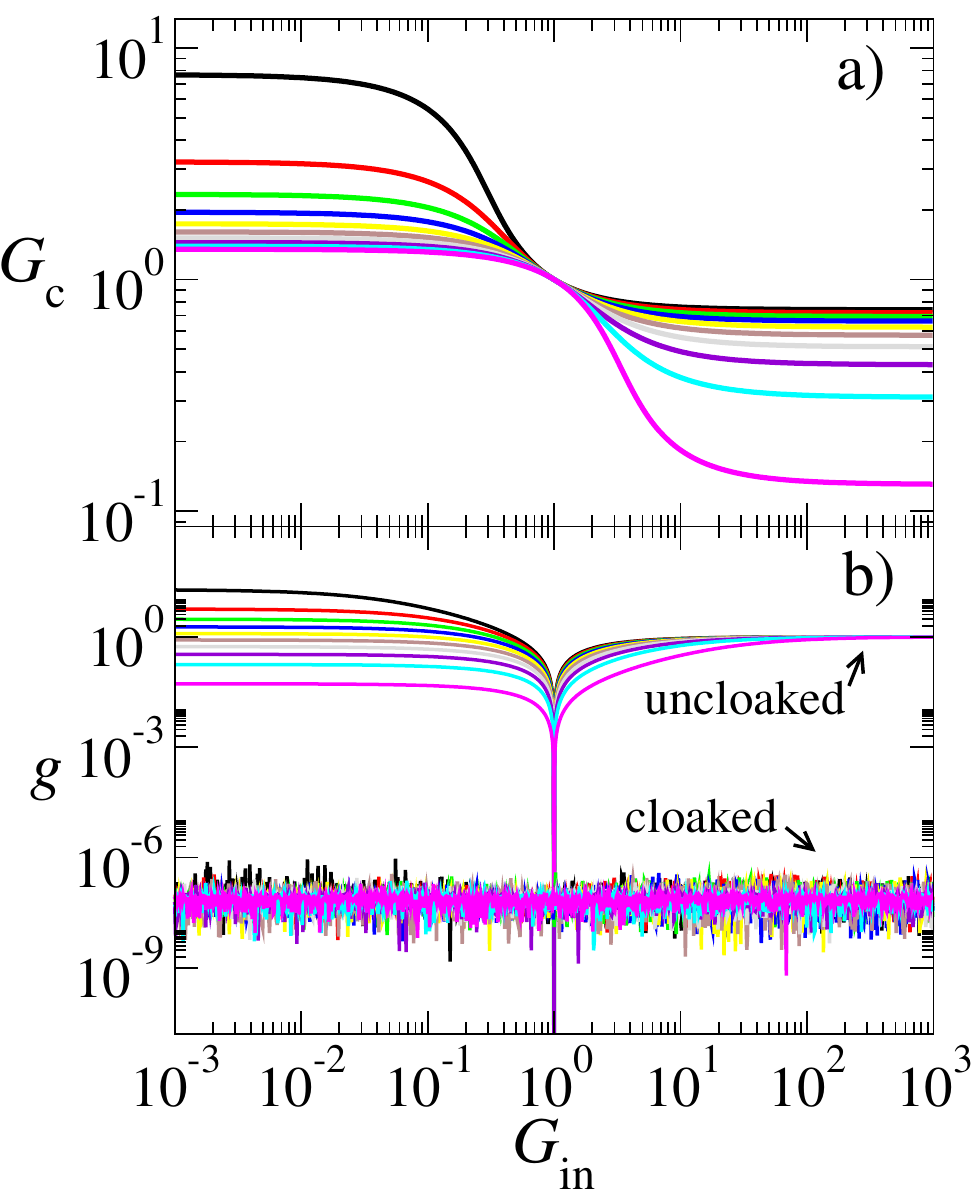} 
\end{center}
  \caption{Cloaking compression by surrounding an inclusion of radius $\Rin=1$ with a single annulus extending from $R=\Rin$ to $\Rc=2$.  {\bf a)} Cloak modulus $\Gc$ as a function of the inclusion modulus.  The Poisson ratio $\nu$ is uniform across the whole system, with $\nu=0.9, 0.7, 0.5, 0.3, 0.1, -0.1, -0.3, -0.5, -0.7, -0.9$ in curves black, red, green, blue, yellow, brown, grey, violet, cyan and magenta top to bottom at the left. {\bf b)} Cloaking performance evidenced by comparing the degree of disturbance to the displacement field (Eqn.~\ref{eqn:ga}) for a cloaked  inclusion (bottom curves) with that for an uncloaked inclusion (top curves). Colour scheme as in a).} 
  \label{fig:compression} 
\end{figure}

\subsubsection{Cloaking bulk compression}
\label{sec:compression}

\begin{figure}[!t]
  \begin{center}
           \includegraphics[width=0.9\columnwidth]{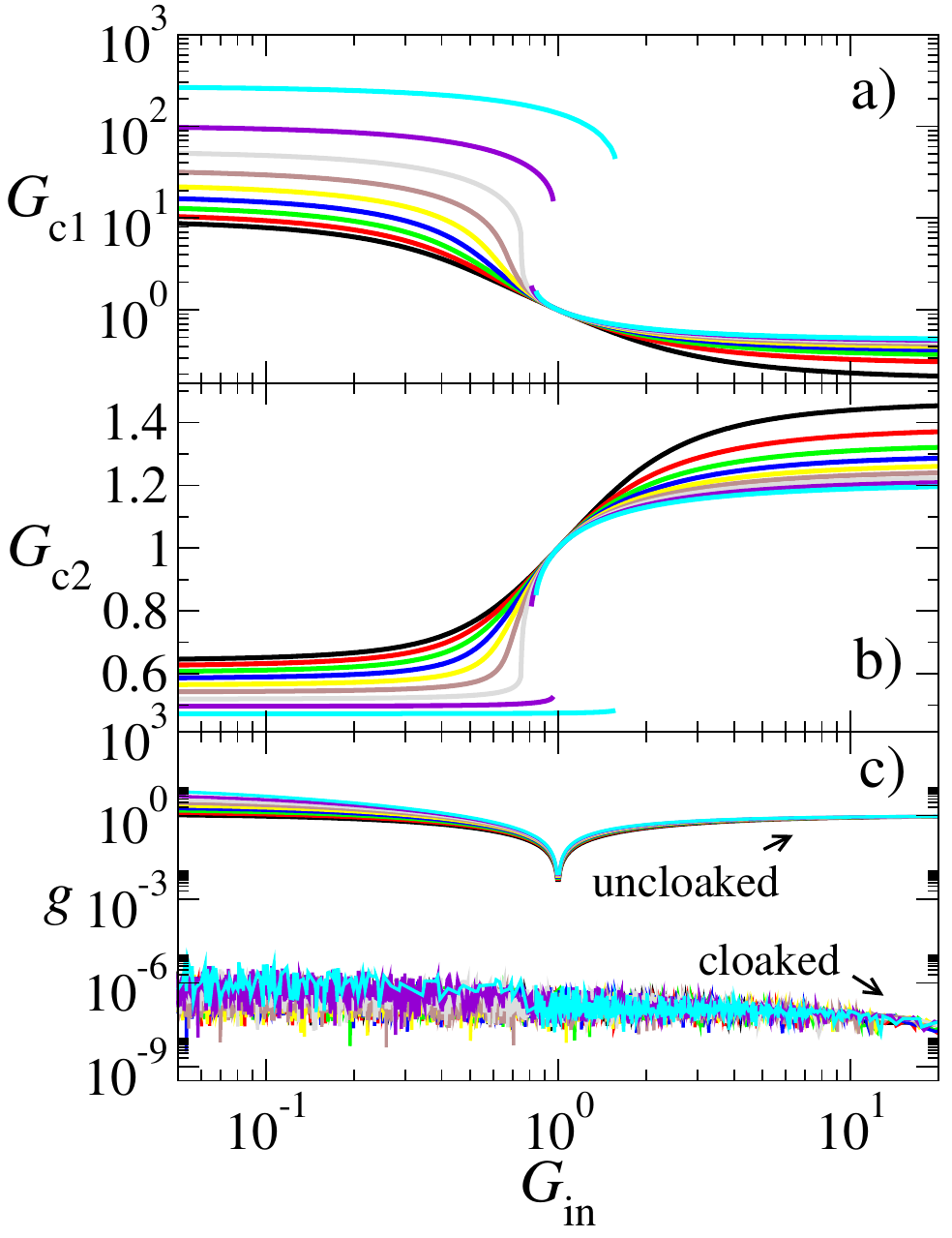} 
  \end{center}
  \caption{Cloaking arbitrarily mixed simple and pure shear by surrounding an inclusion of radius $\Rin=1$ with two concentric annuli, with the inner annulus extending from $\Rin=1$ to $R=3/2$ and the outer annulus  from  $R=3/2$ to $\Rc=2$. {\bf a + b)} Modulus $G_{\rm c1}$ and $G_{\rm c2}$ of the inner and outer cloak annulus respectively, as a function of the inclusion modulus.  The Poisson ratio $\nu$ is uniform across the system, with $\nu=0.9, 0.7, 0.5, 0.3, 0.1, -0.1, -0.3, -0.5, -0.7$ in curves black, red, green, blue, yellow, brown, grey, violet, cyan bottom to top for $G_{\rm c1}$ and top to bottom for $G_{\rm c2}$. {\bf c)} Cloaking performance evidenced as in Fig.~\ref{fig:compression}b).} 
  \label{fig:shear} 
\end{figure}

As a warmup to cloaking an arbitrary admixture of compression and shear, we consider first the simpler task of cloaking only compression $\dV$ (or expansion $\dV<0$). Here the shear components $\dg=\de=0$ and subsystems III) and IV)  have the trivial solution $\uvec=0$, giving  $|u_{\theta s-1}|=|u_{\theta s-3}|=|u_{\theta c-1}|=|u_{\theta c-3}|=0$: the deformation field in isotropic compression is independent of $\theta$, as expected.

Our cloaking metric in Eqn.~\ref{eqn:ga} thus reduces to $g=|u_{r-1}|$, and the task of cloaking to engineering $u_{r-1}=0$ in the surrounding medium. Recall that this quantity appears in the solution vector $\uvec$ of subsystem II), the governing matrix
 $\atens$ of which encodes  Eqns.~\ref{eqn:interface1} and~\ref{eqn:interface3}, and so depends on  the elastic moduli  $\Gin$, $\Gcn$  and $\Go=1$ and Poisson ratios $\nuin,\nucn$ and $\nuo$. Indeed, for arbitrary chosen $\Gin,\nuin,\Gcn,\nucn$, this system $\atens\cdot\uvec=\bvec$ will be exactly specified, with each component of its solution $\uvec$ non-zero in general. 

To achieve cloaking, we need a single additional degree of freedom, which can be tuned to engineer the single condition $u_{r-1}=0$ in the surrounding medium.  Accordingly, we need a single cloaking annulus of tuneable modulus $\Gc$. Linearity further ensures that, for any inclusion $\Gin$, the same cloaking $\Gc$ will cloak any compression $\dV$. Numerically solving subsystem II) with a single annulus,  we employ the downhill simplex method~\cite{teukolsky1992numerical} to minimise the surrounding medium's $g=|u_{r -1}|$ in the solution vector $\uvec$ across values of the cloak modulus $\Gc$, given any inclusion modulus $\Gin$. See  Fig.~\ref{fig:compression}.  We achieve a cloaking metric  $g=O(10^{-6})$, limited only by numerical tolerance,  indicating essentially perfect cloaking.

\begin{figure}[!t]
\begin{center}
     \includegraphics[width=0.9\columnwidth]{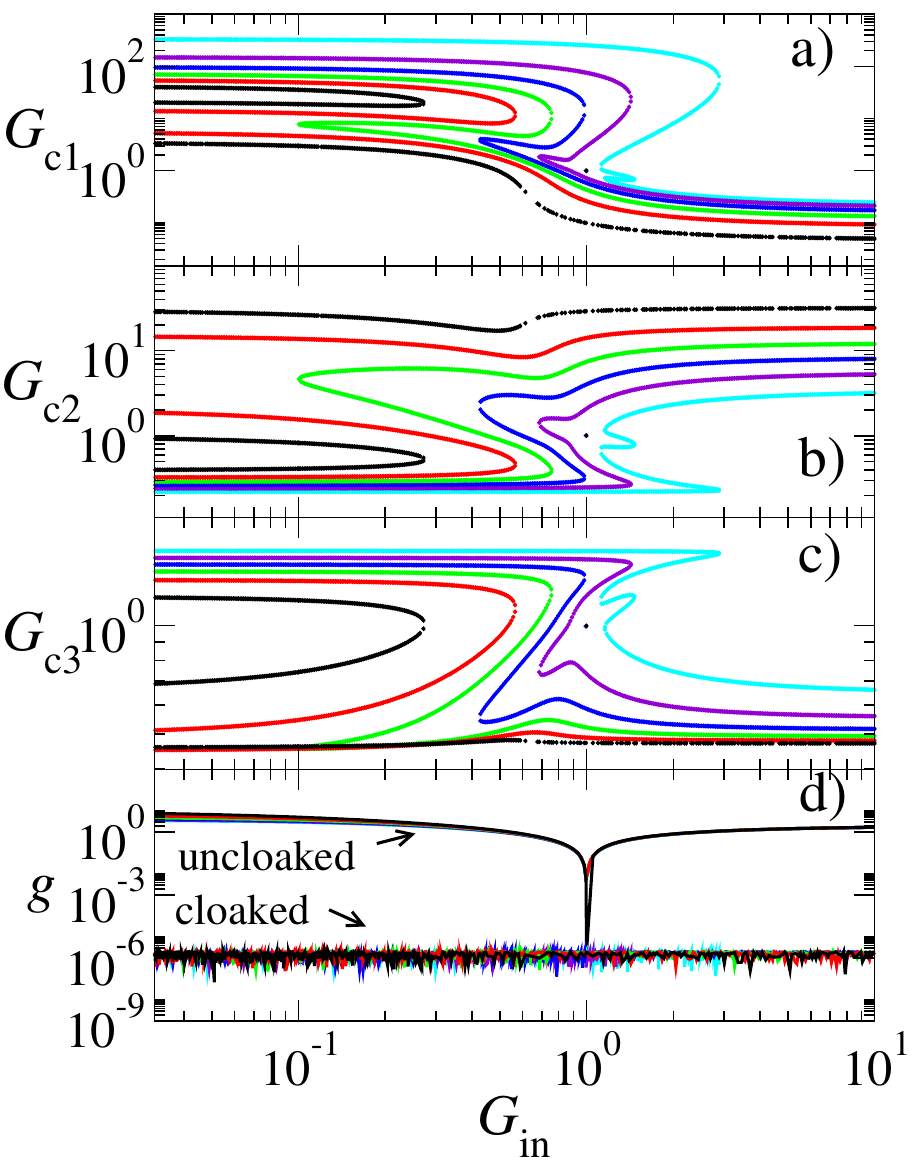} 
\end{center}
  \caption{Cloaking arbitrarily mixed compression, simple shear and pure shear by surrounding an inclusion of radius $\Rin=1$ by three concentric annuli, with the inner annulus extending from  $\Rin=1$ to  $R=4/3$, the middle annulus from $R=4/3$ to $R=5/3$ and the outer annulus from $R=5/3$ to  $\Rc=2$. {\bf a-c)} Modulus $G_{\rm c1}$, $G_{\rm c2}$ and $G_{\rm c3}$  of the inner, middle and outer  cloak annulus, as a function of the  inclusion modulus. The Poisson ratio $\nu$ is uniform across the system in all cases, with $\nu=0.8, 0.7, 0.6, 0.5, 0.4, 0.3$ in curves black, red, green, blue, violet, cyan bottom to top at the right for $G_{\rm c1}$. {\bf d)} Cloaking performance evidenced as in Fig.~\ref{fig:compression}b).} 
  \label{fig:mixed} 
\end{figure}

\subsubsection{Cloaking arbitrarily mixed pure shear and simple shear}
\label{sec:shear}

We now turn to cloaking shear, for now with zero compression $\dV=0$. Here subsystem II  has the trivial solution $\uvec=0$ giving  $|u_{r-1}|=0$ in the surrounding medium. Our cloaking metric thus reduces to 
$g=|u_{\theta s-1}|+|u_{\theta s-3}|+|u_{\theta c-1}|+|u_{\theta c-3}|$ in which $u_{\theta c-1},u_{\theta c-3}$ appear in the solution vector of subsystem IV) and $u_{\theta s-1},u_{\theta s-3}$ in that of III). Recall that III) and IV) are however the same, so that  eliminating $u_{\theta c-1}$ and $u_{\theta c-3}$  automatically also eliminates $u_{\theta s-1}$ and $u_{\theta s-3}$. The task of cloaking thus reduces to engineering the twin conditions $u_{\theta c-1}=0$ and $u_{\theta c-3}=0$ in the surrounding medium. 

As usual, given arbitrary $\Gin, \Gcn$ in the governing matrix $\atens$ of IV), each component of its solution vector $\uvec$ will be non-zero in general.  To achieve cloaking, we need two additional degrees of freedom, tuneable to engineer the two conditions $u_{\theta c-1}=0$ and $u_{\theta c-3}=0$. Accordingly,  we now need two cloaking annuli with tuneable moduli $G_{\rm c1}$ and $G_{\rm c2}$. Given linearity and the equivalence of III) and IV), the same cloaking $G_{\rm c1}$ and $G_{\rm c2}$ will then cloak any arbitrary admixture of pure shear $\de$ and simple shear $\dg$, for any given inclusion modulus $\Gin$. Numerically solving subsystem III) with two annuli,  we  minimise $|u_{\theta c-1}| +|u_{\theta c-3}|$ in its solution vector $\uvec$ across values of  $G_{\rm c1}$ and $G_{\rm c2}$, for any given inclusion modulus $\Gin$. The results are shown  Fig.~\ref{fig:compression}, which again confirms essentially perfect cloaking, with  $g=O(10^{-6})$.

\subsubsection{Cloaking arbitrarily mixed compression and shear}
\label{sec:mixed}

We consider finally cloaking shear and compression combined. Here we must consider both subsystems II) and IV)=III), aiming to reduce our metric $g=|u_{r-1}|+|u_{\theta c-1}|+|u_{\theta c-3}|$ to zero. (We have removed from Eqn.~\ref{eqn:ga} the quantity $|u_{\theta -1}|$, which relates to solid body rotation and is automatically zero as noted above. We have also removed $|u_{\theta s-1}|,|u_{\theta s-3}|$, which are automatically zero once $|u_{\theta c-1}|,|u_{\theta c-3}|$ are eliminated, as also noted above.) We accordingly now require three tuneable quantities -- three cloaking annuli of separately tuneable moduli $G_{\rm c1}, G_{\rm c2}$ and $G_{\rm c3}$ -- in order that each of $|u_{r-1}|$, $|u_{\theta c-1}|$ and $|u_{\theta c-3}|$ can be tuned to zero.  Linearity of the governing equations then means that the same combination of $G_{\rm c1}, G_{\rm c2}$ and $G_{\rm c3}$ will cloak any arbitrary admixture of compression and shear, for any given inclusion modulus $\Gin$. Numerically solving subsystems I) and IV)=III) simultaneously with three annuli, we minimise $g=|u_{r-1}|+|u_{\theta c-1}|+|u_{\theta c-3}|$ in the solution across $G_{\rm c1}, G_{\rm c2}$ and $G_{\rm c3}$  for any $\Gin$. The results are shown in Fig.~\ref{fig:mixed}, again showing essentially perfect cloaking. 

We have only however been able to find cloaking solutions under mixed compression and shear in Fig.~\ref{fig:mixed} for conventional materials of positive Poisson ratio $\nu$ and not auxetics with $\nu<0$. Recall that we did however achieve cloaking in auxetics under compression and shear separately in Figs.~\ref{fig:compression} and~\ref{fig:shear}.

\begin{figure}[!t]
\begin{center}
      \includegraphics[width=0.9\columnwidth]{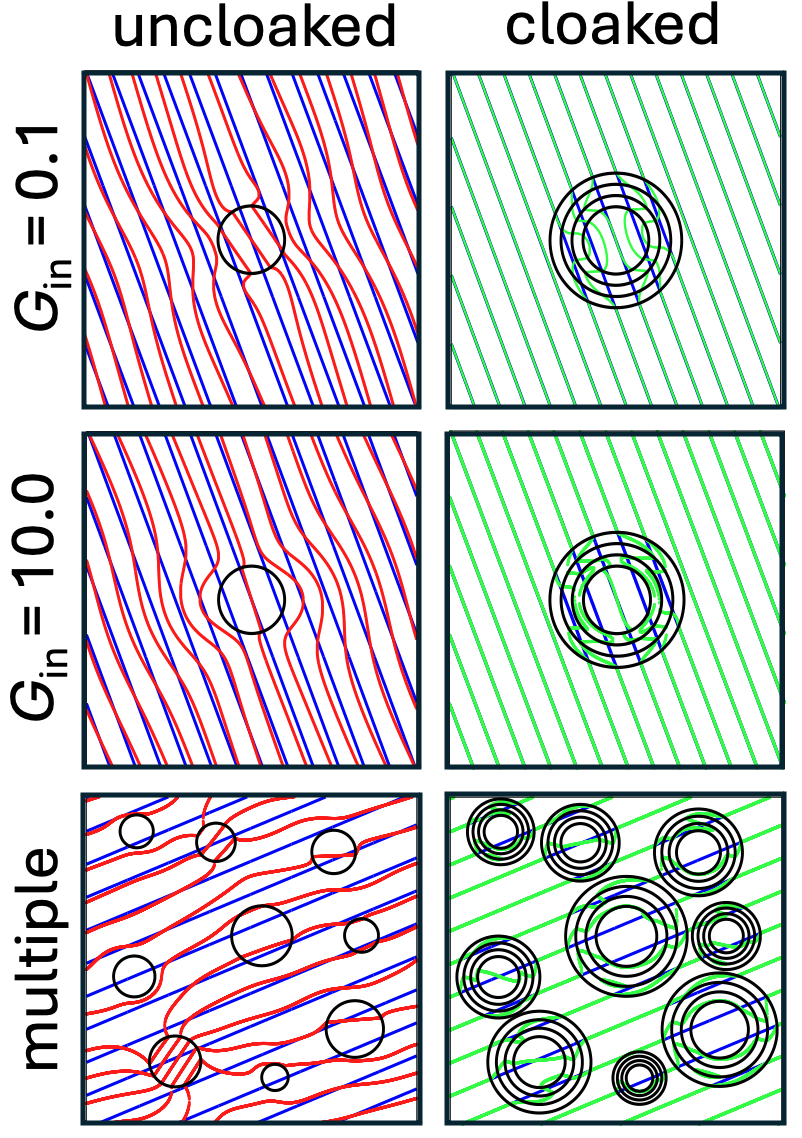}
\end{center}
  \caption{Direct numerical simulation in a biperiodic box with contour lines showing constant horizontal displacement.  {\bf Left:} uncloaked. {\bf Right:}  cloaked by three  annuli. Blue lines: affine deformation field without any inclusion. Red lines: deformation field with inclusion but without cloak. Green lines: deformation field with inclusion and cloak, almost indistinguishable outside cloak from affine  field. {\bf Top:} single inclusion of modulus $\Gin=0.1$ and cloak moduli $G_{\rm c1}=18.83, G_{\rm c2}=0.83, G_{\rm c3}=0.74$ outwards. {\bf Middle:} single inclusion of modulus $\Gin=10.0$ and cloak moduli $G_{\rm c1}=0.050, G_{\rm c2}=31.54, G_{\rm c3}=0.47$. {\bf Bottom:} nine inclusions of random radii placed randomly. $\log_{10}\Gin$ for each inclusion is chosen randomly from a top hat distribution between -0.1 and +1.0. For each inclusion the cloaking moduli $G_{\rm c1}, G_{\rm c2}$ and $G_{\rm c3}$ were taken from the analytical predictions in Fig.~\ref{fig:mixed}. Each deformation component $\dV,\de,\dg$ is chosen randomly from a top hat distribution between $-1.00$ and $+1.00$, with $\dV=-0.18,\de=0.83,\dg=0.49$ for the single inclusion  and   $\dV=0.45,\de=-0.24,\dg=-0.99$ for the multiple inclusions.  Poisson ratio $\nu=0.8$ for surrounding material, cloak and inclusion.  }
  %cloaking3multiple.c_NI9_dV0.45_de-%0.24_dg-0.99_nu0.8_aspect1.0_R0.075_alpha1.0_rf10.3333333333_rf20.6666666666_eta1.0_Lx1.0_Ly1.0_Nx2048_Ny2048_Dt0.003_threshold1.0e-6  }
  \label{fig:snapshot} 
\end{figure}

\section{Confirmation by direct numerical simulation}
\label{sec:numerics}

So far, we have performed analytical calculations to predict conditions for perfect cloaking. We now confirm  these predictions by direct numerical simulation of an elastic material in a biperiodic box. On a $d=2$ dimensional  lattice of $N\times N$ sites, we numerically solve Eqns.~\ref{eqn:strain} to~\ref{eqn:forceBalance} as follows. First, we add to the stress in Eqn.~\ref{eqn:stress1} a dissipative contribution of viscosity $\eta$, $\sigma_{ij}(r_i,t)\to \sigma_{ij}(r_i,t)+2\eta D_{ij}$, with strain rate  
$D_{ij}=\dot{\epsilon}_{ij}=\tfrac{1}{2}\left(\partial_iv_j+\partial_jv_i\right)$ and velocity $v_i=\dot{u}_i$. We then use a time-stepping algorithm, with two separate sub-steps at each timestep. The first comprises an elastic update, in which the strain at each site is incremented $\epsilon_{ij}\to \epsilon_{ij} + \delta t D_{ij}$, with  timestep $\delta t$. This updated strain $\epsilon_{ij}$ is then used to calculate the updated elastic stress $\sigma_{ij}$ in Eqn.~\ref{eqn:stress1}. In the second substep, this elastic stress is transformed into Fourier space with periodic boundary conditions. Imposing force balance on the total stress field, including the dissipative contribution, then allows the non-affine contribution to the velocity and thence the velocity gradient field to be calculated. This is added to the homogeneous affine imposed strain rate to give the total strain rate, which is used again in the first substep at the next timestep. This process is repeated until the spatially averaged magnitude of the strain rate tensor falls below a small threshold $O(10^{-6})$, indicating that static equilibrium has been attained.

We simulate first a single inclusion of low or high modulus $\Gin=0.1$ or $\Gin=10.0$, in each case subject to a randomly chosen admixture of compression and shear, with three cloaking annuli with moduli taken from our analytical predictions in Fig.~\ref{fig:mixed}. The results are shown in the top two rows of Fig.~\ref{fig:snapshot}, confirming excellent cloaking. 

\begin{figure}[!t]
    \includegraphics[width=\columnwidth]{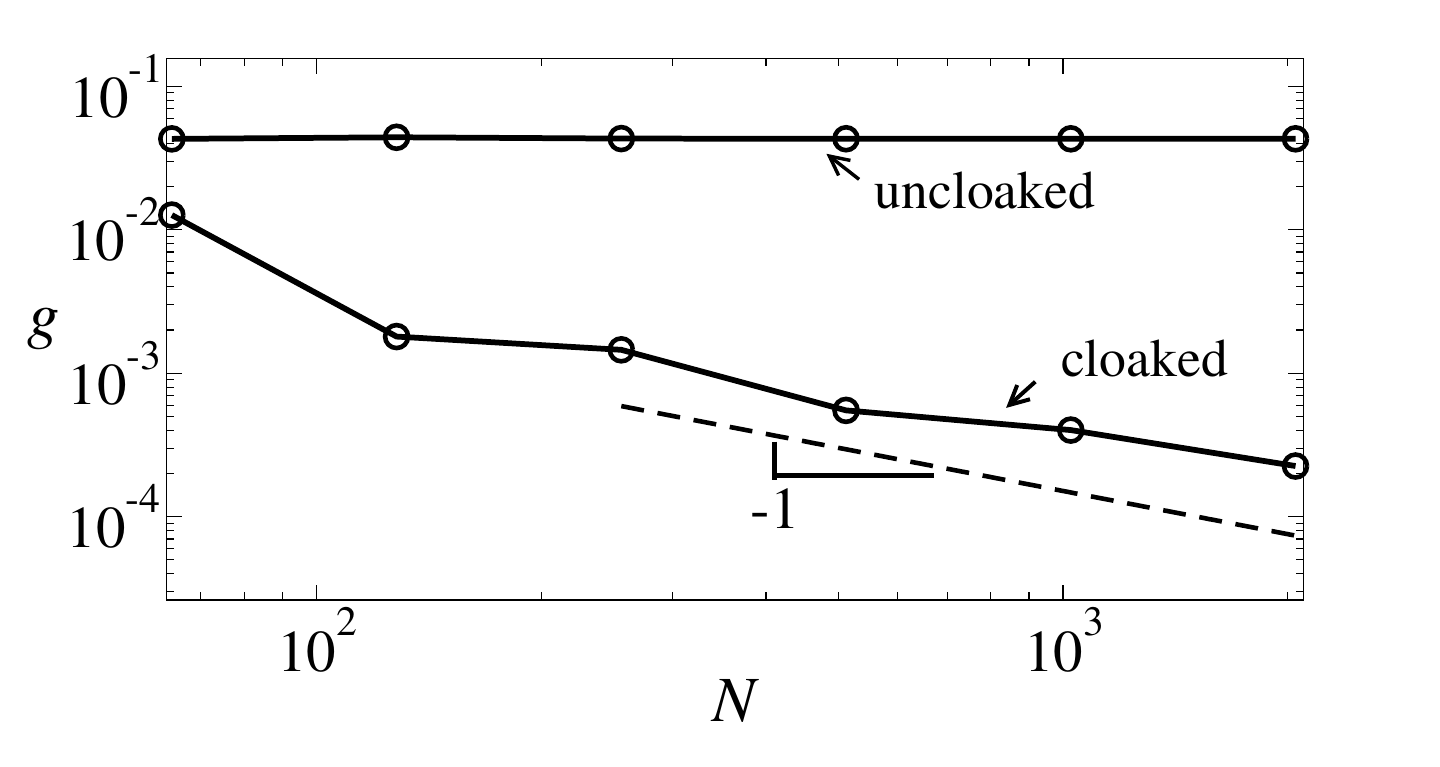} 
  \caption{Cloaking performance as a function of number of numerical grid points in direct numerical simulation, evidenced by comparing the degree of disturbance to the displacement field (Eqn.~\ref{eqn:ga}) for a cloaked void or inclusion (bottom) with that for an uncloaked void or inclusion (top). Inclusion modulus $\Gin=10.0$. Cloaking moduli $G_{\rm c1}=0.050, G_{\rm c2}=31.54, G_{\rm c3}=0.47$ taken from the analytical predictions in Fig.~\ref{fig:mixed}. The applied deformation $\dV=-0.18,\de=0.83,\dg=0.49$. Poisson ratio $\nu=0.8$ for surrounding material, cloak and inclusion.  } 
  \label{fig:converge} 
\end{figure}

In our analytical calculation, we assumed perfectly sharp interfaces between annuli. In our numerics, the interfaces have a small but non-zero thickness set by the mesh size. This causes a small correction to perfect cloaking, which we characterise in our numerics by the metric $g$, defined now as the integral over space outside the cloaked inclusion of the squared deviation of the deformation field from the affine homogeneous one. This is indeed non-zero (Fig.~\ref{fig:converge}) but decreases with increasing mesh resolution, suggesting that perfect cloaking will be recovered for perfectly sharp interfaces as $N\to\infty$, consistent with our analytical considerations above.

Given that a single inclusion can be cloaked in the way we have described, even at near field close to the cloaking perimeter, it follows that multiple inclusions  arranged with arbitrarily high packing fraction in a surrounding medium can also be cloaked. This is confirmed by our simulation results in the bottom row of Fig.~\ref{fig:snapshot}. This indicates a possible route to fabricating composite materials with any density of  inclusions in a surrounding medium, however high, in which the strain and stress fields remain uniform outwith the inclusions. This may facilitate the design of composites with the same overall bulk mechanical properties as an originally homogeneous medium, but with (for example) a  lower density overall. It may also potentially mitigate the increased risk of cracking that could arise from stress concentrations in a composite without cloaking.

\begin{figure}[!t]
\begin{center}
      \includegraphics[width=0.95\columnwidth]{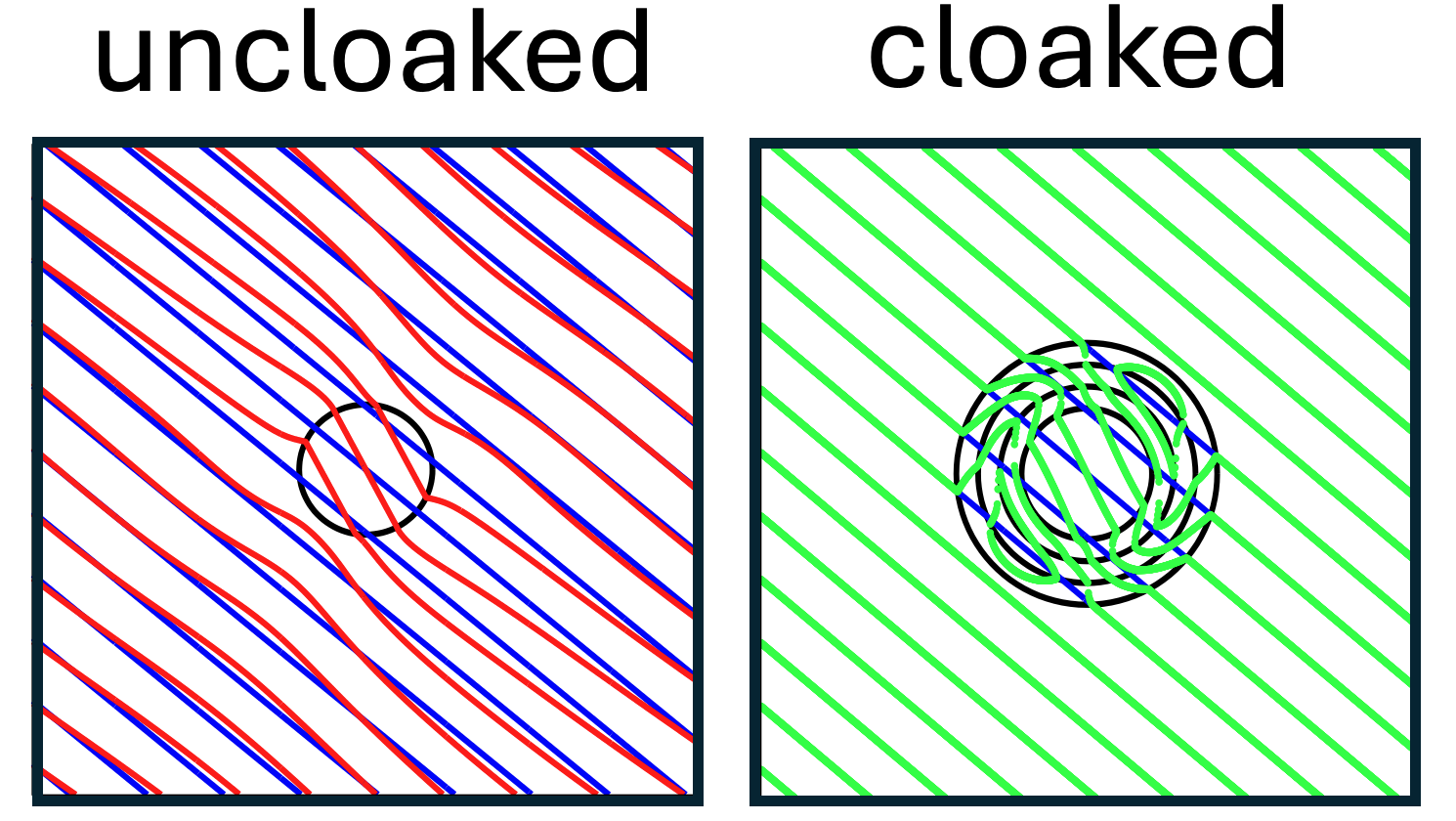} 
\end{center}
  \caption{Direct numerical simulation in a biperiodic box, with contour lines showing constant horizontal displacement. {\bf Left:} uncloaked. {\bf Right:} cloaked by three annuli. Blue lines: affine deformation field without any inclusion. Red lines: deformation field with inclusion but without cloak. Green lines: deformation field with inclusion and cloak, almost indistinguishable outside cloak from affine  field. Here we chose each Poisson ratio $\nuin=0.699, \nucone=0.990, \nuctwo =0.587 , \nucthree=0.266 , \nuo=0.570$  randomly from a top hat distribution between 0.0 and 1.0, $\log_{10} \Gin$ randomly from a top hat distribution between $-1.0$ and 1.0, giving $\Gin = 1.54$, and each of $\dV=-0.44, \de=0.57, \dg=0.31$ randomly from a top hat distribution between $-1.0$ and 1.0.  The shear modulus of each annulus was then tuned to give cloaking: $\Gcone=0.971, \Gctwo=0.417,\Gcthree=1.598$. } 
  \label{fig:Poisson} 
\end{figure}

\section{Discussion}
\label{sec:conclusion}

We have shown theoretically that essentially perfect elastostatic  cloaking of a circular inclusion in a homogeneous surrounding medium can be achieved by means of a cloak comprising three concentric annuli, in any arbitrary combination of compression and shear.  A full range of circular inclusions can be cloaked in this way, from soft  to stiff. In consequence,  we suggest also that an inclusion of any arbitrary shape can also be cloaked, by first enveloping it in a stiff circle, then cloaking the combined structure with three annuli as described. Compared with cloaks fabricated from metamaterials comprising many subunits, the cloaks suggested here are simple  and significantly more effective. 

Indeed, in achieving essentially perfect cloaking performance, the strain and so also stress fields outside the cloaked inclusion are undisturbed compared with those in a homogeneous medium without an inclusion. Accordingly, our results should apply equally to arbitrary imposed loads as well as the arbitrary imposed strains that we have considered here. 

Our analytical calculations in Sec.~\ref{sec:analytics}  showed that respectively one, two and three tuneable cloaking shear moduli are needed to cloak compression, shear, and arbitrarily mixed compression and shear respectively. For simplicity,  we performed our numerical calculations  in the associated Figs.~\ref{fig:compression},~\ref{fig:shear} and~\ref{fig:mixed} assuming a uniform Poisson ratio $\nu=\nuin=\nucn=\nuo$ across the entire device (inclusion, all cloaking annuli and the surrounding material). Our analytical approach does not however place any requirement on  the value of the Poisson ratio  being precisely tuned in any part of the device. As noted above, this  is important for fabricating a device experimentally, because  precisely prescribing both the shear modulus and Poisson ratio of any given material component  is essentially impossible to achieve in practice.

With this motivation, we show finally in Fig.~\ref{fig:Poisson} a fully cloaked state for an arbitrarily imposed deformation  for a given inclusion and surrounding material  in which the Poisson ratio of each cloaking annulus $\nucn$ was  chosen randomly from a flat distribution between zero and one. For this given, arbitrarily chosen set of values of $\nucn$, the moduli $\Gcn$ of the cloaking annuli were then tuned to ensure cloaking. We emphasize, therefore, that for any given combination of materials of (linear elastic) inclusion and surrounding medium, an experimentalist is free (first) to choose the Poisson ratio of each cloaking annulus arbitrarily. For any such set of Poisson ratios, thus chosen, it then suffices only to precisely select the shear modulus of each annulus according to the prescription given above.

Our calculations have predicted perfect cloaking in the limit of perfectly sharp interfaces between annuli. Any real interface will in practice have a small but finite thickness set by microscopics, suggesting slightly imperfect cloaking. It is worth noting, however, that imperfect interfaces have also been suggested as a way to enhance cloaking~\cite{bertoldi2007structural,bigoni1998asymptotic,he20023d,ru1998interface,wang2012neutrality}.

We have focused  on circular inclusions in $d=2$ spatial dimensions. Future work should extend the analytical arguments developed here to spherical inclusions in $d=3$ dimensions. Our calculations have furthermore been performed in the  linear elastic regime, and therefore apply in systems subject to small deformation and/or loads that are small on scale of material's modulus. We defer to future work a study of cloaking in the nonlinear elastic regime.

{\it Acknowledgements ---}  This project has received funding from the European Research Council (ERC) under the European Union's Horizon 2020 research and innovation programme (grant agreement No. 885146).

%\bibliography{cloaking.bib}


\end{document}